\documentclass[pre,twocolumn,showpacs,preprintnumbers,amsmath,amssymb,groupedaddress,superscriptaddress]
{revtex4}

\usepackage{graphicx}% Include figure files
\usepackage{dcolumn}% Align table columns on decimal point
\usepackage{bm}% bold math
\usepackage{psfrag}
\usepackage{epsfig}
\usepackage{amsmath}
\usepackage{amssymb}
\usepackage{color}
\usepackage{fancyheadings}
\usepackage{mathbbol}

\newcommand{\nn}{\nonumber}
\newcommand{\bra}{\langle}
\newcommand{\ve}{\vert}
\newcommand{\ket}{\rangle}

\begin{document}

%\preprint{APS/123-QED}

\title{Local Versus Global Thermal States: Correlations and the Existence of Local Temperatures}

\author{Michael Hartmann} 
\email{michael.hartmann@dlr.de}
\affiliation{Institute of Technical Physics, DLR Stuttgart}
\affiliation{Institute of Theoretical Physics I, University of Stuttgart}
\author{G\"unter Mahler}
\affiliation{Institute of Theoretical Physics I, University of Stuttgart}
\author{Ortwin Hess}
\affiliation{Advanced Technology Institute, University of Surrey}

\date{\today}

\begin{abstract}
We consider a quantum system consisting of a regular chain of elementary subsystems with nearest
neighbor interactions and assume that the total system is in a canonical state with temperature $T$.
We analyze under what condition the state factors into a product of canonical density matrices
with respect to groups of $n$ subsystems each, and when these groups have the same temperature $T$.
While in classical mechanics the validity of this procedure only depends on the size of the groups $n$,
in quantum mechanics the minimum group size $n_{min}$ also depends on the temperature $T \,$!
As examples, we apply our analysis to a harmonic chain and different types of Ising spin chains.
We discuss various features that show up due to the characteristics of the models considered.
For the harmonic chain, which successfully describes thermal properties of insulating solids,
our approach gives a first quantitative estimate of the minimal length scale on which temperature can exist:
This length scale is found to be constant for temperatures above the Debye temperature
and proportional to $T^{-3}$ below.
\end{abstract}

\maketitle

% ---------------------------------------------------------------------------
%

\section{Introduction}

Thermodynamics is among the most successfully and extensively applied theoretical concepts in
physics. Notwithstanding, the various limits of its applicability are not fully understood
\cite{GemmerOtte2001,Allahverdyan2000}.

Of particular interest is its microscopic limit.
Down to which length scales can its standard concepts meaningfully be defined and employed?

Besides its general importance, this question has become increasingly relevant recently since
amazing progress in the synthesis and processing of materials with structures on
nanometer length scales has created a demand for better understanding of thermal properties of
nanoscale devices, individual nanostructures and nanostructured materials
\cite{Cahill2003,Williams1986,Varesi1998,Schwab2000}.
Experimental techniques have improved to such an extent that the measurement of thermodynamic
quantities like temperature with a spatial resolution on the nanometer scale seems within reach
\cite{Gao2002,Pothier1997,Aumentado2001}. 

To provide a basis for the interpretation of present day and future experiments in nanoscale physics
and technology and to obtain a better understanding of the limits of thermodynamics,
it is thus indispensable to clarify the applicability of thermodynamical concepts
on small length scales starting from the most fundamental theory at hand, i. e. quantum mechanics.
In this context, one question appears to be particularly important and interesting:
Can temperature be meaningfully defined on nanometer length scales?

The existence of thermodynamical quantities, i. e. the existence of the thermodynamic limit strongly
depends on the correlations between the considered parts of a system.

With increasing size, the volume of a region in space grows faster than its surface.
Thus effective interactions between two regions, provided they are short ranged, become less relevant
as the sizes of the regions increase.
This scaling behavior is used to show that correlations between a region and its environment
become negligible in the limit of infinite region size and that therefore the thermodynamic limit exists
\cite{Fisher1964,Ruelle1969,Lebowitz1969}.

To explore the minimal region size needed for the application of thermodynamical concepts, situations far
away from the thermodynamic limit should be analyzed. On the other hand, effective correlations between
the considered parts need to be small enough \cite{Schmidt1998,Hartmann2003a}.

The scaling of interactions between parts of a system compared to the energy contained in the parts
themselves thus sets a minimal length scale on which correlations are
still small enough to permit the definition of local temperatures.
It is the aim of this paper to study this connection quantitatively

Some attempts to generalize thermodynamics such that it applies to small systems have been made
\cite{Hill1994,Hill2001,Rajagopal2004}.
These approaches consider ensembles of independent, i.e. noninteracting, small systems.
By introducing an additional thermodynamical potential they take into account the
surface effects of the small systems. However, since the interactions between the small systems are
neglected, these concepts cannot capture the physics of the correlations.
This shortcoming is also obvious from the results:
The correction terms they predict do not depend on temperature, whereas it is well known, that correlations
become more important the lower the temperature.

Recently the impact of quantum correlations, i. e. entanglement on macroscopic properties
of solids and phase transitions has drawn considerable attention \cite{Osterloh2002,Roscilde2004,Vedral2003}.
Since our analysis of criteria for local temperatures is based on a study of correlations,
our theoretical approach is a promising tool to provide further insight into
the role of correlations in solid state physics.

We adopt here the convention that a local temperature exists if the considered part of the system is in a
canonical state, where the distribution is an exponentially decaying function of energy characterized by one
single parameter. This implies that there is a one-to-one mapping between temperature and
the expectation values of observables, by which temperature is usually measured. Temperature measurements
based on different observables will thus yield the same result, contrary to distributions with several
parameters. In large systems composed of very many subsystems, the density of states is a strongly growing
function of energy \cite{Tolman1967}. If the distribution were not exponentially decaying,
the product of the density of states times the distribution would not have a pronounced peak
and thus physical quantities like energy would not have ``sharp'' values.

There have been atempts to describe systems which are not in an equilibrium state but in some sense close to it
with a generalized form of thermodynamics, that has additional system parameters.
Such a situation appears for example in glasses \cite{Nieuwenhuizen1998}.
Our approach analyzes whether thermodynamics in its standard form can apply locally. A study whether a
generalized form of thermodynamics might apply even more locally should be a subject of future research.

A typical setup where the minimal length scale we calculate becomes relevant could be the measurement of a
temperature profile with very high resolution etc.
One is thus interested in scenarios where the entire sample is expected to be in a stationary state.
In most cases this state is close to a thermal equilibrium state \cite{Kubo1985}.

Based on the above arguments and noting that a quantum description becomes imperative at nanoscopic scales,
the following approach appears to be reasonable:
Consider a large homogeneous quantum system, brought into a thermal state via interaction with its environment,
divide this system into subgroups and analyze for what
subgroup-size the concept of temperature is still applicable.

Harmonic lattice models are a standard tool for the description of thermal properties of solids.
We therefore apply our theory to a harmonic chain model to get estimates that are expected to be relevant
for real materials and might be tested by experiments.

Recently, spin chains have been subject of extensive studies in condensed matter physics and
quantum information theory.
Thus correlations and possible local temperatures in spin chains are of interest, both
from a theoretical and experimental point of view \cite{Wang2002,Kenzelmann2002}.
We study spin chains with respect to our present purpose and compare their characteristics with the harmonic chain.

This paper is organized as follows: In section \ref{general}, we present the general
theoretical approach which derives two conditions on the effective group interactions and the global
temperature.
In the following two sections we apply the general consideration to two concrete models and derive
estimates for the minimal subgroup size.
Section \ref{harmonicchain} deals with a harmonic chain, a model with an infinite energy spectrum.
In contrast, a spin chain has a bounded energy spectrum. Section \ref{isingchain}
therefore discusses an Ising spin chain in a transverse field. In the conclusions section \ref{conclusion},
we compare the results for the different models considered and indicate further interesting topics. 
%
%-------------------------------------------------------------------------------------------------------
%XXXXXXXXXXXXXXXXXXXXXXXXXXXXXXXXXXXXXXXXXXXXXXXXXXXXXXXXXXXXXXXXXXXXXXXXXXXXXXXXXXXXXXXXXXXXXXXXXXXXXXX
%XXXXXXXXXXXXXXXXXXXXXXXXXXXXXXXXXXXXXXXXXXXXXXXXXXXXXXXXXXXXXXXXXXXXXXXXXXXXXXXXXXXXXXXXXXXXXXXXXXXXXXX
%-------------------------------------------------------------------------------------------------------
%
\section{General Theory\label{general}} 

We consider a homogeneous (i.e. translation invariant) chain of elementary quantum
subsystems with nearest neighbor interactions.
The Hamiltonian of our system is thus of the form \cite{Mahler1998},
\begin{equation}\label{hamil}
H = \sum_{i} H_i + I_{i,i+1}
\end{equation}
where the index $i$ labels the elementary subsystems. $H_i$ is the Hamiltonian of subsystem $i$
and $I_{i,i+1}$ the interaction between subsystem $i$ and $i+1$.
We assume periodic boundary conditions.

We now form $N_G$ groups of $n$ subsystems each
(index $i \rightarrow (\mu-1) n + j; \: \mu = 1, \dots, N_G; \: j = 1, \dots, n$)
and split this Hamiltonian into two parts,
\begin{equation}
\label{hsep}
H = H_0 + I,
\end{equation}
where $H_0$ is the sum of the Hamiltonians of the isolated groups,
\begin{eqnarray}\label{isogroups}
H_0 & = & \sum_{\mu=1}^{N_G} \left( \mathcal{H}_{\mu} - I_{\mu n,\mu n + 1} \right) \enspace \enspace
\textrm{with} \nn \\
\mathcal{H}_{\mu} & = & \sum_{j=1}^n H_{n (\mu - 1) + j} + I_{n (\mu-1) + j,\, n (\mu-1) + j + 1} 
\end{eqnarray}
and $I$ contains the interaction terms of each group with its neighbor group,
\begin{equation}
I = \sum_{\mu=1}^{N_G} I_{\mu n,\mu n + 1}.
\end{equation}
We label the eigenstates of the total Hamiltonian $H$
and their energies with the Greek indices $(\varphi, \psi)$ and eigenstates and energies
of the group Hamiltonian  $H_0$ with Latin indices $(a, b)$,
\begin{equation}
\label{prodstate}
H \: \ve \varphi \ket = E_{\varphi} \: \ve \varphi \ket \enspace \enspace \textrm{and} \enspace \enspace
H_0 \: \ve a \ket = E_a \: \ve a \ket.
\end{equation}
Here, the states $\ve a \ket$ are products of group eigenstates
\begin{equation}
\ve a \ket = \prod_{\mu = 1}^{N_G} \ve a_{\mu} \ket,
\end{equation}
where
$\left( \mathcal{H}_{\mu} - I_{\mu n, \mu n + 1} \right) \ve a_{\mu} \ket = E_{\mu} \ve a_{\mu} \ket$.
$E_{\mu}$ is the energy of one subgroup only and $E_a = \sum_{\mu=1}^{N_G} E_{\mu}$.

\subsection{Thermal State in the Product Basis}
We assume that the total system is in a thermal state with the density matrix
\begin{equation}
\label{candens}
\bra \varphi \ve \hat \rho \ve \psi \ket = \frac{e^{- \beta E_{\varphi}}}{Z} \: \delta_{\varphi \psi}
\end{equation}
in the eigenbasis of $H$. Here, $Z$ is the partition sum and $\beta = (k_B T)^{-1}$
the inverse temperature with Boltzmann's constant $k_B$ and temperature $T$.
Transforming the density matrix (\ref{candens}) into the eigenbasis of $H_0$ we obtain
\begin{equation}
\label{newrho}
\bra a \ve \hat \rho \ve a \ket =
\int_{E_0}^{E_1} w_a (E) \: \frac{e^{- \beta E}}{Z} \: dE
\end{equation}
for the diagonal elements in the new basis. Here, the sum over all states $\ve \varphi \ket$ has been replaced
by an integral over the energy. $E_0$ is the energy of the ground state and $E_1$ the upper limit of the
spectrum. For systems with an energy spectrum that does not have an upper bound, the limit
$E_1 \rightarrow \infty$ should be taken. The density of conditional probabilities $w_a (E)$ is given by
\begin{equation}
w_a (E) = \frac{1}{\Delta E} \,
\sum_{\{ \ve \varphi \ket: E \le E_{\varphi} < E + \Delta E \}}
\ve \bra a \ve \varphi \ket \ve^2
\end{equation}
where $\Delta E$ is small and the sum runs over all states $\ve \varphi \ket$ with eigenvalues
$E_{\varphi}$ in the interval $[E,E + \Delta E)$.
To compute the integral of equation (\ref{newrho}) we need to know the distribution of the
conditional probabilities $w_a (E)$. 

The state $\ve a \ket$ is not an eigenstate of the total Hamiltonian $H$.
Thus, if $H$ would be measured in the state $\ve a \ket$, eigenvalues of $H$ would be obtained with
certain probabilities: $w_a (E)$ is the density of this probability distribution.
Since the hamiltonian $H$ is the sum of hamiltonians of the groups, the situation has some
analogies to a sum of random variables. This indicates that there might exist a central limit theorem
for the present quantum system, provided the number of groups becomes very large \cite{Billingsley1995} .
Since the state $\ve a \ket$ is not translation invariant and since $H$ also contains the group
interactions, the central limit theorem has to be of a Lyapunov (or Lindeberg)
type for mixing sequences \cite{Linnik1971}.
One can indeed show that such a quantum central limit theorem exists for the present model
\cite{Hartmann2003,Hartmann2004b} and that $w_a (E)$ thus converges to a Gaussian normal distribution
in the limit of infinite number of groups $N_G$,
\begin{equation}
\label{gaussian_dist}
\lim_{N_G \to \infty} w_a (E) = \frac{1}{\sqrt{2 \pi} \Delta_a}
\exp \left(- \frac{\left(E - E_a - \varepsilon_a \right)^2}{2 \, \Delta_a^2} \right),
\end{equation}
where the quantities $\varepsilon_a$ and $\Delta_a$ are defined by 
\begin{eqnarray}
\varepsilon_a & \equiv & \bra a \ve H \ve a \ket - E_a \\
\Delta_a^2 & \equiv & \bra a \ve H^2 \ve a \ket - \bra a \ve H \ve a \ket^2.
\end{eqnarray}
$\varepsilon_a$ is the difference between the energy expectation value of the distribution $w_a (E)$ and the
energy $E_a$, while $\Delta_a^2$ is the variance of the energy $E$ for the distribution $w_a (E)$.
Note that $\varepsilon_a$ has a classical counterpart while $\Delta_a^2$ is purely quantum mechanical.
It appears because the commutator $[H,H_0]$ is nonzero, and the distribution $w_a(E)$ therefore has nonzero
width. The two quantities $\varepsilon_a$ and $\Delta_a^2$ can also be expressed in terms of the
interaction only (see eq. (\ref{hsep})),
\begin{eqnarray}
\varepsilon_a & = & \bra a \ve I \ve a \ket\\
\Delta_a^2 & = & \bra a \ve I^2 \ve a \ket - \bra a \ve I \ve a \ket^2,
\end{eqnarray}
meaning that $\varepsilon_a$ is the expectation value and $\Delta_a^2$ the squared width of the interactions
in the state $\ve a \ket$.

The rigorous proof of equation (\ref{gaussian_dist}) is given in \cite{Hartmann2003} and based on
the following two assumptions:
The energy of each group $\mathcal{H}_{\mu}$ as defined in equation (\ref{isogroups}) is bounded, i. e.
\begin{equation}
\label{bounded}
\bra \chi \ve \mathcal{H}_{\mu} \ve \chi \ket \le C
\end{equation}
for all normalized states $\ve \chi \ket$ and some constant $C$, and
\begin{equation}
\label{vacuumfluc}
\bra a \ve H^2 \ve a \ket - \bra a \ve H \ve a \ket^2 \ge N_G \, C'
\end{equation}
for some constant $C' > 0$.

In scenarios where the energy spectrum of each elementary subsystem has an upper limit, such as spins,
condition (\ref{bounded}) is met a priori.
For subsystems with an infinite energy spectrum, such as harmonic oscillators,
we restrict our analysis to states where the energy of every group,
including the interactions with its neighbors, is bounded. Thus, our considerations do not apply
to product states $\ve a \ket$, for which all the energy was located in only one group or only a small
number of groups. Since $N_G  \gg 1$, the number of such states is vanishingly small compared
to the number of all product states.

If conditions (\ref{bounded}) and (\ref{vacuumfluc}) are met, equation (\ref{newrho})
can be computed for $N_G  \gg 1$ \cite{Hartmann2004}:
\begin{equation}
\label{newrho2}
\begin{split}
\bra a \ve \hat \rho \ve a \ket = \frac{1}{2 \, Z} \,
\exp \left(- \beta y_a + \frac{\beta^2 \Delta_a^2}{2} \right) \hspace{2.6cm} \\
\left[\textrm{erfc} \left( \frac{E_0 - y_a + \beta \Delta_a^2}{\sqrt{2} \, \Delta_a} \right) -
\textrm{erfc} \left( \frac{E_1 - y_a + \beta \Delta_a^2}{\sqrt{2} \, \Delta_a} \right) \right]
\end{split}
\end{equation}
where $y_a = E_{a} + \varepsilon_a$ and $\textrm{erfc} (x)$ is the conjugate Gaussian error function,
\begin{equation}
\label{def_errorf}
\textrm{erfc}(x) = \frac{2}{\sqrt{\pi}} \, \int_x^{\infty} e^{- s^2} \, ds.
\end{equation}
The second error function in (\ref{newrho2}) only appears if the energy is bounded and the
integration extends from the energy of the ground state $E_0$ to the upper limit of the spectrum $E_1$.

Note that $y_a$ is a sum of $N_G$ terms and that $\Delta_a$ fulfills equation (\ref{vacuumfluc}).
The arguments of the conjugate error functions thus grow proportional to $\sqrt{N_G}$ or stronger.
If these arguments divided by $\sqrt{N_G}$ are finite (different from zero),
the asymptotic expansion of the error function \cite{Abramowitz1970} may thus be used for $N_G \gg 1$:
\begin{equation}
\label{asymptotic_errorf}
\textrm{erfc}(x) \approx \left\{
\begin{array}{lcl}
{\displaystyle \frac{\exp \left(- x^2 \right)}{\sqrt{\pi} \, x}}  & \textrm{for} & x \rightarrow \infty \\
{\displaystyle 2 + \frac{\exp \left(- x^2 \right)}{\sqrt{\pi} \, x}} & \textrm{for} & x \rightarrow - \infty 
\end{array}
\right.
\end{equation}
Inserting this approximation into equation (\ref{newrho2}) and using $E_0 < y_a < E_1$ shows
that the second conjugate error function, which contains the upper limit of the energy spectrum,
can always be neglected compared to the first, which contains the ground state energy.

The same type of arguments show that the normalizations of the Gaussian in equation (\ref{gaussian_dist})
is correct although the energy range does not extend over the entire real axis ($ -\infty, \infty$).

Applying the asymptotic expansion (\ref{asymptotic_errorf}), equation (\ref{newrho2}) can be taken to read
\begin{equation}
\label{newrho_lower}
\bra a \ve \hat \rho \ve a \ket = \frac{1}{Z}
\exp \left[- \beta \left(E_{a} + \varepsilon_a - \frac{\beta \Delta_a^2}{2} \right) \right]
\end{equation}
for
$\left(E_0 - E_{a} - \varepsilon_a + \beta \Delta_a^2 \right) / \left( \sqrt{2 N_G} \, \Delta_a \right) < 0$
and
\begin{equation} 
\label{newrho_greater}
\bra a \ve \hat \rho \ve a \ket =
\frac{ {\displaystyle \exp \left(- \beta E_0 - \frac{(E_a + \varepsilon_a - E_0)^2}{2 \Delta_a^2} \right)}}
{{\displaystyle \sqrt{2 \pi} \: Z \: \frac{E_0 - E_a - \varepsilon_a + \beta \Delta_a^2}{\Delta_a}}},
\end{equation}
for
$\left(E_0 - E_{a} - \varepsilon_a + \beta \Delta_a^2 \right) / \left( \sqrt{2 N_G} \, \Delta_a \right) > 0$.

The off diagonal elements $\bra a \ve \hat \rho \ve b \ket$ vanish for\linebreak 
$\ve E_a - E_b \ve > \Delta_a + \Delta_b$ because the overlap of the two distributions of conditional
probabilities becomes negligible.\linebreak For $\ve E_a - E_b \ve < \Delta_a + \Delta_b$, the transformation
involves an integral over frequencies and thus these terms are significantly smaller than
the entries on the diagonal.

\subsection{Conditions for Local Thermal States}
We now test under what conditions the density matrix $\hat \rho$ may be approximated by a product
of canonical density matrices with temperature $\beta_{loc}$ for each subgroup $\mu = 1, 2, \dots, N_G$.
Since the trace of a matrix is invariant under basis transformations, it is sufficient to verify
the correct energy dependence of the product density matrix.
If we assume periodic boundary conditions,
all reduced density matrices are equal and their product
is of the form $\bra a \ve \hat \rho \ve a \ket \propto \exp(- \beta_{loc} E_a)$.
We thus have to verify whether the logarithm of rhs of equations (\ref{newrho_lower}) and
(\ref{newrho_greater}) is a linear function of the energy $E_a$, 
\begin{equation} \label{log}
\ln \left( \bra a \ve \hat \rho \ve a \ket \right) \approx - \beta_{loc} \, E_a + c,
\end{equation}
where $\beta_{loc}$ and $c$ are constants.

Note that equation (\ref{log}) does not imply that the
occupation probability of an eigenstate $\ve \varphi \ket$ with energy $E_{\varphi}$ and a product state with
the same energy $E_a \approx E_{\varphi}$ are equal. Since $\beta_{loc}$ and $\beta$ enter into the
exponents of the respective canonical distributions, the difference between both has significant consequences
for the occupation probabilities; even if $\beta_{loc}$ and $\beta$ are equal with very
high accuracy, but not exactly the same, occupation probabilities may differ by several orders of magnitude,
provided that the energy range is large enough. 

We exclude negative temperatures ($\beta > 0$).
Equation (\ref{log}) can only be true for
\begin{equation} \label{cond_const}
\frac{E_a + \varepsilon_a  - E_0}{\sqrt{N_G} \, \Delta_a} > \beta \frac{\Delta_a^2}{\sqrt{N_G} \, \Delta_a} ,
\end{equation}
as can be seen from equations (\ref{newrho_lower}) and (\ref{newrho_greater}).
In this case, $\bra a \ve \hat \rho \ve a \ket$ is given by (\ref{newrho_lower}) and to satisfy (\ref{log}),
$\varepsilon_a$ and $\Delta_a^2$ furthermore have to be of the following form:
\begin{equation}
\label{cond_linear_1} 
- \varepsilon_a + \frac{\beta}{2} \, \Delta_a^2 \approx  c_1 E_a + c_2 
\end{equation}
where $c_1$ and $c_2$ are constants.
Note that $\varepsilon_a$ and $\Delta_a^2$ need not be functions of $E_a$ and therefore in general
cannot be expanded in a Taylor series.

To ensure that the density matrix of each subgroup $\mu$ is approximately canonical, one needs to satisfy
(\ref{cond_linear_1}) for each subgroup $\mu$ separately;
\begin{equation}
\label{cond_linear_2} 
- \frac{\varepsilon_{\mu - 1} + \varepsilon_{\mu}}{2} + \frac{\beta}{4} \,
\left(\Delta_{\mu - 1}^2 + \Delta_{\mu}^2 \right)
+ \frac{\beta}{6} \, \tilde{\Delta}_{\mu}^2 \, \approx \, c_1  \, E_{\mu} + c_2
\end{equation}
where $\varepsilon_{\mu} = \bra a \ve I_{\mu n, \mu n + 1} \ve a \ket$
with $\varepsilon_a = \sum_{\mu=1}^{N_G} \varepsilon_{\mu}$,\linebreak
$\Delta_{\mu}^2 = \bra a \ve \mathcal{H}_{\mu}^2 \ve a \ket -
\bra a \ve \mathcal{H}_{\mu} \ve a \ket^2$ and
$\tilde{\Delta}_{\mu}^2 =
\sum_{\nu = \mu-1}^{\mu+1} \bra a \ve \mathcal{H}_{\nu-1} \mathcal{H}_{\nu} +
\mathcal{H}_{\nu} \mathcal{H}_{\nu-1} \ve a \ket -
2 \bra a \ve \mathcal{H}_{\nu-1} \ve a \ket \bra a \ve \mathcal{H}_{\nu} \ve a \ket$ and

Temperature becomes intensive, if the constant $c_1$ vanishes,
\begin{equation} \label{intensivity}
\left| c_1 \right| \ll 1 \enspace \enspace \Rightarrow \enspace \enspace \beta_{loc} = \beta.
\end{equation}
If this was not the case, temperature would not be intensive, although it might exist locally.

\begin{figure}[h]
\centering
\psfrag{eta}{\small \hspace{-0.35cm} \raisebox{0.55cm}{$\eta (E)$}}
\psfrag{pb}{\small \hspace{0.7cm} \raisebox{0.05cm}{$\bra \varphi \ve \hat \rho \ve \varphi \ket$}}
\psfrag{prod}{\small \hspace{0.08cm} \raisebox{0.1cm}{$\eta (E) \cdot \bra \varphi \ve \hat \rho \ve \varphi \ket$}}
\psfrag{E1}{\small \hspace{-0.48cm} $E_{min}$}
\psfrag{E2}{\small \hspace{-0.23cm} $E_{max}$}
\psfrag{E}{\small \hspace{-0.45cm} \raisebox{-0.7cm}{$\overline{E}$}}
\psfrag{n}{}
\psfrag{c1}{$\: E$}
\epsfig{file=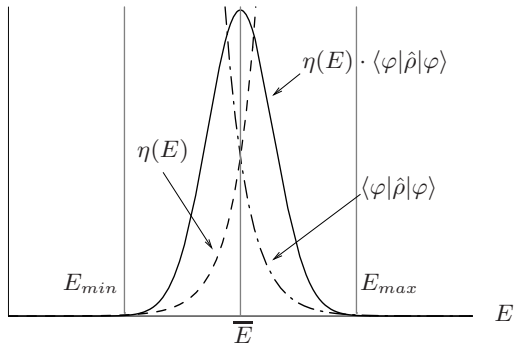,width=7cm}
\caption{The product of the density of states $\eta (E)$ times the occupation probabilities
$\bra \varphi \ve \hat \rho \ve \varphi \ket$ forms a strongly pronounced peak at $E = \overline{E}$.}
\label{peak}
\end{figure}

It is sufficient to satisfy conditions (\ref{cond_const}) and (\ref{cond_linear_2}) for an adequate energy
range $E_{min} \le E_{\mu} \le E_{max}$ only.
For large systems with a modular structure, i.e. a system composed of a large number of
subsystems, the density of states is
typically a rapidly growing function of energy \cite{GemmerPhD, Tolman1967}. If the total system is in a
thermal state, occupation probabilities decay exponentially with energy. The product of these two
functions is thus sharply peaked at the expectation value of the energy $\overline{E}$ of the total
system $\overline{E} + E_0 = $Tr$(H \hat \rho)$, with $E_0$ being the ground state energy
(see figure \ref{peak}).
%
% ---------------------------------------------------------------------------
%
% Figure Peak
%
The energy range thus needs to be centered around this peak and large enough.
On the other hand it must not be larger than the range of values $E_{\mu}$ can take on.
Therefore a pertinent and ``safe'' choice for $E_{min}$ and $E_{max}$ is
\begin{equation} \label{e_range}
\begin{array}{rcl}
E_{min} & = & \textrm{max}
\left( \left[E_{\mu}\right]_{min} \, , \,
\frac{1}{\alpha} \frac{\overline{E}}{N_G} + \frac{E_0}{N_G} \right)\\
E_{max} & = & \textrm{min}
\left( \left[E_{\mu}\right]_{max} \, , \, \alpha
\frac{\overline{E}}{N_G} + \frac{E_0}{N_G} \right)
\end{array}
\end{equation}
where $\alpha \gg 1$ and $\overline{E}$ will in general depend on the global temperature.
In equation (\ref{e_range}), $\left[E_{\mu}\right]_{min}$ and $\left[E_{\mu}\right]_{max}$ denote
the minimal and maximal values $E_{\mu}$ can take on.

\begin{figure}[h]
\psfrag{3}{\small \hspace{+0.2cm} $ $}
\psfrag{4}{\small \hspace{+0.2cm} $ $}
\psfrag{5}{\small \raisebox{-0.1cm}{%$5$
$ $}}
\psfrag{6}{\small \raisebox{-0.1cm}{%$6$
$ $}}
\psfrag{7}{\small \raisebox{-0.1cm}{%$7$
$ $}}
\psfrag{8}{\small \raisebox{-0.1cm}{%$8$
$ $}}
\psfrag{E1}{\small \hspace{-0.48cm} $E_{low}$}
\psfrag{E2}{\small \hspace{-0.25cm} $E_{high}$}
\psfrag{n}{\hspace{-0.3cm} \raisebox{0.1cm}{$\ln \left( \bra a \ve \hat \rho \ve a \ket \right)$}}
\psfrag{c1}{$\: E$}
\epsfig{file=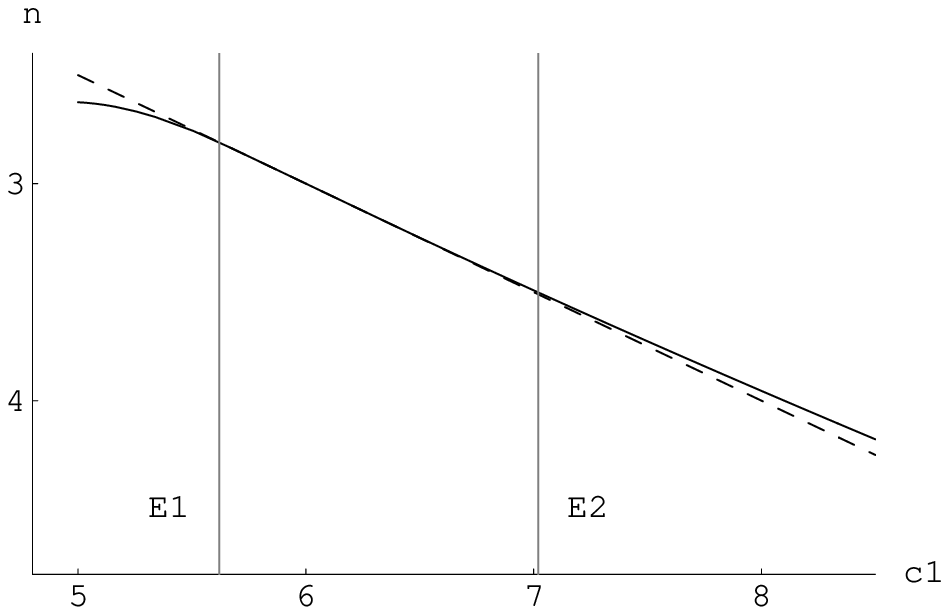,width=7cm}
\caption{$\ln \left( \bra a \ve \hat \rho \ve a \ket \right)$ for $\hat \rho$ as in equation (\ref{newrho2}) (solid line)
and a canonical density matrix $\hat \rho$ (dashed line) for a harmonic chain.}
\label{visual}
\end{figure}

Figure \ref{visual} shows the logarithm of equation (\ref{newrho2}) and the logarithm of a
canonical distribution with the same $\beta$ for a harmonic chain as an example. The actual
density matrix is more mixed than the canonical one.
In the interval between the two vertical lines, both criteria
(\ref{cond_const}) and (\ref{cond_linear_2}) are satisfied.
For $E < E_{low}$ (\ref{cond_const}) is violated and (\ref{cond_linear_2}) for $E > E_{high}$. To allow
for a description by means of canonical density matrices, the
group size needs to be chosen such that $E_{low} < E_{min}$ and $E_{high} > E_{max}$.
%
% ---------------------------------------------------------------------------
%
% Figure Visual
%

For a model obeying equations (\ref{bounded}) and (\ref{vacuumfluc}), the two conditions
(\ref{cond_const}) and (\ref{cond_linear_2}), which constitute the general result of this article,
must both be satisfied. These fundamental criteria will now be applied to some concrete examples.
%
%-------------------------------------------------------------------------------------------------------
%XXXXXXXXXXXXXXXXXXXXXXXXXXXXXXXXXXXXXXXXXXXXXXXXXXXXXXXXXXXXXXXXXXXXXXXXXXXXXXXXXXXXXXXXXXXXXXXXXXXXXXX
%XXXXXXXXXXXXXXXXXXXXXXXXXXXXXXXXXXXXXXXXXXXXXXXXXXXXXXXXXXXXXXXXXXXXXXXXXXXXXXXXXXXXXXXXXXXXXXXXXXXXXXX
%-------------------------------------------------------------------------------------------------------
%
\section{Harmonic Chain\label{harmonicchain}} 

As a representative for the class of systems with an infinite energy spectrum,
we consider a harmonic chain of $N_G \cdot n$ particles of mass $m$ and
spring constant $\sqrt{m} \, \omega_0$. In this case, the Hamiltonian reads
\begin{eqnarray}
H_i & = & \frac{m}{2} \, p_i^2 + \frac{m}{2} \, \omega_0^2 \, q_{i}^2 \\
I_{i, i+1} & = & - m \, \omega_0^2 \, q_{i} \, q_{i+1},
\end{eqnarray}
where $p_i$ is the momentum of the particle at site $i$ and $q_{i}$ the displacement from its equilibrium
position $i \cdot a_0$ with $a_0$ being the distance between neighboring particles at equilibrium.
We divide the chain into $N_G$ groups of $n$ particles each and thus get
a partition of the type considered above.

The Hamiltonian of one group is diagonalized by a Fourier transform and the definition of creation
and annihilation operators $a_{k}^{\dagger}$ and $a_{k}$ for the Fourier modes (see
appendix \ref{diagon_harmonic_chain}).
\begin{equation}
E_a = \sum_{\mu=1}^{N_G} \sum_{k} \omega_k \left( n_k^a (\mu) + \frac{1}{2} \right),
\end{equation}
where $k = \pi l / (a_0 \, (n+1))$ $(l = 1, 2, \dots, n)$ and the frequencies $\omega_k$ are given by
$\omega^2_{k} = 4 \, \omega_0^2 \, \sin^2(k a / 2)$.
$n_k^a (\mu)$ is the occupation number of mode $k$ of
group $\mu$ in the state $\ve a \ket$. We chose units, where $\hbar = 1$.

We first verify that the harmonic chain model fulfills the conditions for the applicability of the quantum
central limit theorem (\ref{gaussian_dist}).
To see that it satisfies the condition (\ref{vacuumfluc}) one needs to
express the group interaction $I_{\mu n, \mu n + 1}$ in terms of
$a_{k}^{\dagger}$ and $a_{k}$, which yields $\tilde{\Delta}_{\mu} = 0$ for all $\mu$
and therefore
\begin{equation}
\Delta_a^2 = \sum_{\mu=1}^{N_G} \Delta_{\mu}^2,
\end{equation}
where $\Delta_{\mu}$, the width of one group interaction, reads
\begin{equation}
\label{harmsigma}
\begin{split}
\Delta^2_{\mu} = \left( \frac{2}{n+1} \right)^2
\left( \sum_{k} \cos^2 \left( \frac{k a_0}{2} \right) \, \omega_{k} \,
\left( n_{k} + \frac{1}{2} \right) \right) \cdot \\
\cdot \left( \sum_{p} \cos^2 \left( \frac{p a_0}{2} \right) \, \omega_{p} \,
\left( m_{p} + \frac{1}{2} \right) \right).
\end{split}
\end{equation}
$\Delta^2_{\mu}$ has a minimum value since all $n_{k} \ge 0$ and all $m_{p} \ge 0$.
In equation (\ref{harmsigma}), $k$ labels the modes of group $\mu$ with occupation numbers $n_k$
and $p$ the modes of group $\mu+1$ with occupation numbers $m_p$.
The width $\Delta^2_a$ thus fulfills condition (\ref{vacuumfluc}).

Since the spectrum of every single oscillator is infinite, condition (\ref{bounded}) can only
be satisfied for states, for which the energy of the system is distributed among a substantial fraction
of the groups, as discussed in section \ref{general}.

We now turn to analyze the two criteria (\ref{cond_const}) and (\ref{cond_linear_2}).
The expectation values of the group interactions vanish, $\varepsilon_{\mu} = 0$, while the widths
$\Delta^2_{\mu}$ depend on the occupation numbers $n_{k}$ and therefore on the energies $E_{\mu}$.
We thus have to consider both conditions, (\ref{cond_const}) and (\ref{cond_linear_2}).
To analyze these, we make use of the continuum or Debye approximation \cite{Kittel1983}, requiring
$n \gg 1$, $a_0 \ll l$, where $l = n \, a_0$, and the length of the chain to be finite.
In this case we have $\omega_k = v \, k$ with the constant velocity of sound $v = \omega_0 \, a_0$ and
$\cos (k \, a_0 / 2) \approx 1$. The width of the group interaction thus translates into
\begin{equation}
\label{harmsigma_2}
\Delta^2_{\mu} = \frac{4}{n^2} \, E_{\mu} \, E_{\mu+1}
\end{equation}
where $n+1 \approx n$ has been used. The relevant energy scale is introduced by the thermal expectation
value of the entire chain
\begin{equation}
\label{intenergy}
\overline{E} = N_G n k_B \Theta \left( \frac{T}{\Theta} \right)^2
\int_0^{\Theta / T} \frac{x}{e^x - 1} \, dx,
\end{equation}
and the ground state energy is given by
\begin{equation}
\label{groundenergy}
E_0 = N_G n k_B \Theta \left( \frac{T}{\Theta} \right)^2
\int_0^{\Theta / T} \frac{x}{2} \, dx = \frac{N_G n k_B \Theta}{4}
\end{equation}
We first consider the criterion (\ref{cond_const}).

For a given $E_a = \sum_{\mu} E_{\mu}$, the squared width $\Delta^2_{\mu}$ is largest if all
$E_{\mu}$ are equal, $E_{\mu} = \tilde{E} \: \forall \mu$. Thus (\ref{cond_const}) is hardest to satisfy for
that case, where it reduces to
\begin{equation}
\label{quadratc_ineq}
\tilde{E} - \frac{E_0}{N_G} - \frac{4 \beta}{n^2} \tilde{E}^2 > 0.
\end{equation}
Equation (\ref{quadratc_ineq}) sets a lower bound on $n$. For temperatures where $\overline{E} < E_0$, this
bound is strongest for low energies $\tilde{E}$, while at $\overline{E} > E_0$ it is strongest for high
energies $\tilde{E}$.
Since condition (\ref{cond_linear_2}) is a stronger criterion than condition (\ref{cond_const})
for $\overline{E} > E_0$, we only consider (\ref{quadratc_ineq}) at temperatures where $\overline{E} < E_0$.
In this range, (\ref{quadratc_ineq}) is hardest to satisfy for low energies, i.e. at
$\tilde{E} = (\overline{E} / \alpha  N_G) + (E_0 / N_G)$, where it reduces to
\begin{equation}
\label{harmcond2}
n > \frac{\Theta}{T} \, \frac{\alpha}{4 \overline{e}} \, \left(\frac{4 \overline{e}}{\alpha} + 1 \right)^2,
\end{equation}
with $\overline{e} = \overline{E} / (n N_G k_B \Theta)$.

To test condition (\ref{cond_linear_2}) we take the derivative with respect to $E_{\mu}$ on both sides,
\begin{equation}
\label{subcondition2}
\frac{\beta}{n^2} \left( E_{\mu-1} + E_{\mu+1} - 2 \, \frac{E_0}{N_G} \right) +
\frac{2 \beta}{n^2} \frac{E_0}{N_G} \approx c_1
\end{equation}
where we have separated the energy dependent and the constant part in the lhs.
(\ref{subcondition2}) is satisfied if the energy dependent part is much smaller than one,
\begin{equation}
\label{harmsubcond2}
\frac{\beta}{n^2} \left( E_{\mu-1} + E_{\mu+1} - 2 \frac{E_0}{N_G} \right) \le \delta \ll 1.
\end{equation}
This condition is hardest to satisfy for high energies.
Taking $E_{\mu-1}$ and $E_{\mu+1}$ equal to the upper bound in equation (\ref{e_range}), it yields
\begin{equation}
\label{subcondition3}
n > \frac{2 \alpha}{\delta} \, \frac{\Theta}{T} \, \overline{e},
\end{equation}
where the ``accuracy'' parameter $\delta \ll 1$ quantifies the value of the energy dependent part in
(\ref{subcondition2}).

Since the constant part in the lhs of (\ref{subcondition2}) satisfies
\begin{equation}
\frac{2 \beta}{n^2} \frac{E_0}{N_G} < \frac{\sqrt{\delta}}{\alpha} \,
\left(\frac{1}{\sqrt{2}} - \frac{\sqrt{\delta}}{\alpha} \right) \: \ll 1,
\end{equation}
temperature is intensive.

\begin{figure}[h]
\psfrag{-4.1}{\small \raisebox{-0.1cm}{$10^{-4}$}}
\psfrag{-3.1}{\small \raisebox{-0.1cm}{$10^{-3}$}}
\psfrag{-2.1}{\small \raisebox{-0.1cm}{$10^{-2}$}}
\psfrag{-1.1}{\small \raisebox{-0.1cm}{$10^{-1}$}}
\psfrag{1.1}{\small \raisebox{-0.1cm}{$10^{1}$}}
\psfrag{2.1}{\small \raisebox{-0.1cm}{$10^{2}$}}
\psfrag{3.1}{\small \raisebox{-0.1cm}{$10^{3}$}}
\psfrag{4.1}{\small \raisebox{-0.1cm}{$10^{4}$}}
%\psfrag{1}{\small \hspace{+0.2cm} $10^{1}$}
\psfrag{1}{}
\psfrag{2}{\small \hspace{+0.2cm} $10^{2}$}
%\psfrag{3}{\small \hspace{+0.2cm} $10^{3}$}
\psfrag{3}{}
\psfrag{4}{\small \hspace{+0.2cm} $10^{4}$}
%\psfrag{5}{\small \hspace{+0.2cm} $10^{5}$}
\psfrag{5}{}
\psfrag{6}{\small \hspace{+0.2cm} $10^{6}$}
%\psfrag{7}{\small \hspace{+0.2cm} $10^{7}$}
\psfrag{7}{}
\psfrag{8}{\small \hspace{+0.2cm} $10^{8}$}
\psfrag{n}{\raisebox{0.1cm}{$n_{min}$}}
\psfrag{c1}{$\: T / \Theta$}
\epsfig{file=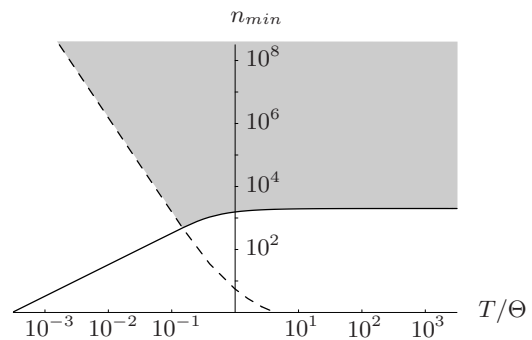,width=7cm}
\caption{Log-log-plot of $n_{min}$ from eq. (\ref{harmcond2}) (dashed line) and
$n_{min}$ from eq. (\ref{subcondition3}) (solid line) for $\alpha = 10$ and $\delta = 0.01$
as a function of $T / \Theta$ for a harmonic chain. $\delta$ and $\alpha$ are defined in equations
(\ref{subcondition3}) and (\ref{e_range}), respectively. Local temperature exists in the shaded area.}
\label{temp}
\end{figure}

Inserting equation (\ref{intenergy}) into equation (\ref{harmcond2}) and (\ref{subcondition3})
one can now calculate the minimal $n$ for given $\delta, \alpha,\Theta$ and $T$.
Figure \ref{temp} shows $n_{min}$ for $\alpha = 10$ and $\delta = 0.01$ given by criterion
(\ref{harmcond2}) and (\ref{subcondition3}) as a function of $T / \Theta$.
Hence, local temperature exists, i. e. local states are canonical for all group sizes larger then
the maximum of the two $n_{min}$-curves plotted in figure \ref{temp}.
%
% ---------------------------------------------------------------------------
%
% Figure Temp1d
%

For high (low) temperatures $n_{min}$ can thus be estimated by
\begin{equation} \label{eq:1}
n_{min} \approx \left\{
\begin{array}{lcr}
2 \, \alpha / \delta & \textrm{for} & T > \Theta\\
\left( 3 \alpha / 2 \pi^2 \right) \, \left( \Theta / T \right)^3 & \textrm{for} & T < \Theta
\end{array}
\right.
\end{equation}

Equation (\ref{eq:1}) also shows the dependence of the results on the ``accuracy parameters''
$\alpha$ and $\delta$. In the whole temperature range, $n_{min} \propto \alpha$, in other words,
the larger one chooses the energy range
where (\ref{cond_const}) and (\ref{cond_linear_2}) should be fulfilled, the larger has to be the number of
particles per group. Furthermore, for high temperatures, $n_{min} \propto \delta^{-1}$, which simply
states that one needs more particles per group to obtain a canonical state with better accuracy.

Since the resulting minimal group sizes $n_{min}$ are larger than $10^3$ for all temperatures,
the application of the Debye approximation is well justified.
%
%-------------------------------------------------------------------------------------------------------
%XXXXXXXXXXXXXXXXXXXXXXXXXXXXXXXXXXXXXXXXXXXXXXXXXXXXXXXXXXXXXXXXXXXXXXXXXXXXXXXXXXXXXXXXXXXXXXXXXXXXXXX
%XXXXXXXXXXXXXXXXXXXXXXXXXXXXXXXXXXXXXXXXXXXXXXXXXXXXXXXXXXXXXXXXXXXXXXXXXXXXXXXXXXXXXXXXXXXXXXXXXXXXXXX
%-------------------------------------------------------------------------------------------------------
%
\section{Ising Spin Chain in a Transverse Field\label{isingchain}} 

In this section we consider an Ising spin chain in a transverse field. For this model the Hamiltonian
reads
\begin{eqnarray} \label{ising_ham}
H_i & = & - B \, \sigma_i^z \nn \\
I_{i, i+1} & = & - \frac{J_x}{2} \, \sigma_i^x \otimes \sigma_{i+1}^x -
\frac{J_y}{2} \, \sigma_i^y \otimes \sigma_{i+1}^y
\end{eqnarray}
where $\sigma_i^x, \sigma_i^y$ and $\sigma_i^z$ are the Pauli matrices. $B$ is the magnetic field
and $J_x$ and $J_y$ are two coupling parameters. We will always assume $B > 0$.

The entire chain with periodic boundary conditions may be diagonalized via successive Jordan-Wigner,
Fourier and Bogoliubov transformations (see appendix \ref{diagon_ising_chain}). The relevant energy scale
is introduced via the thermal expectation value (without the ground state energy)
\begin{equation}
\label{ising_e_quer}
\overline{E} = \frac{n N_G}{2 \pi} \int_{-\pi}^{\pi} dk \,
\frac{\omega_k}{\exp \left( \beta \, \omega_k \right) + 1},
\end{equation}
where $\omega_k$ is given in equation (\ref{ising_frequ}).
The ground state energy $E_0$ is given by
\begin{equation}
\label{ising_e_0}
E_0 = - \frac{n N_G}{2 \pi}
\int_{-\pi}^{\pi} dk \, \frac{\omega_k}{2}.
\end{equation}
Since $N_G \gg 1$, the sums over all modes have been replaced by integrals.

If one partitions the chain into $N_G$ groups of $n$ subsystems each, the groups may also be diagonalized
via a Jordan-Wigner and a Fourier transformation (see appendix \ref{diagon_ising_chain}).
Using the abbreviations  
\begin{equation}
\label{abbreviations}
K = \frac{J_x + J_y}{2 \, B} \enspace \enspace \textrm{and} \enspace \enspace L = \frac{J_x - J_y}{2 \, B},
\end{equation}
the energy $E_a$ reads
\begin{equation}
\label{ising_e_a}
E_a = 2 B \, \sum_{\mu = 1}^{N_G} \sum_{k} \left[ 1 - K \cos(k) \right]
\left(n_k^a(\mu) -\frac{1}{2} \right),
\end{equation}
where $k = \pi l / (n+1)$ ($l = 1, 2, \dots, n$) and $n_k^a(\mu)$ is the fermionic occupation
number of mode $k$ of group $\mu$ in the state $\ve a \ket$. It can take on the values $0$ and $1$.

For the Ising model at hand one has, as for the harmonic chain, $\varepsilon_a = 0$ for all states
$\ve a \ket$, while the squared variance $\Delta_a^2$ reads
\begin{equation}
\Delta_a^2 = \sum_{\mu=1}^{N_G} \Delta_{\mu}^2,
\end{equation}
with
\begin{align}
\label{ising_delta_a}
\Delta_{\mu}^2 & = B^2 \, \left( \frac{K^2}{2} + \frac{L^2}{2} \right) -  \\
- 2 \, B^2 \, \left( K^2 - L^2 \right) &
\left[\frac{2}{n+1} \sum_{k} \sin^2(k) \, \left( n_k^a(\mu) - \frac{1}{2} \right) \right] \cdot \nn \\
\cdot & \left[\frac{2}{n+1} \sum_{p} \sin^2(p) \, \left( n_p^a(\mu+1) - \frac{1}{2} \right) \right] \nn 
\end{align}
where the $n_k^a(\mu)$ are the same fermionic occupation numbers as in equation (\ref{ising_e_a}).

The conditions for the central limit theorem are met for the Ising chain apart from two
exceptions: Condition (\ref{bounded}) is always fulfilled as the Hamiltonian of a single spin has
finite dimension. As follows from equation (\ref{ising_delta_a}), condition (\ref{vacuumfluc})
is satisfied except for one single state in the case where $J_x = J_y$ ($L = 0$) and $J_x = - J_y$ ($K = 0$)
respectively. These two states have $\Delta_{\mu}^2 = 0$ and thus $\Delta_{a}^2 < N_G C'$.
The state for $L = 0$ is the one where all occupation numbers $n_k^a(\mu)$ vanish and the state for
$K = 0$ is the state with alternating occupation numbers
$n_k^a(\mu) = 0$, $n_k^a(\mu + 1) = 1$, $n_k^a(\mu + 2) = 0, \dots$ (for all $k$ each).
As there is, for given parameters, at most one state that does not fulfill (\ref{vacuumfluc}),
the fraction of states where our theory does not apply is negligible for $N_G \gg 1$.

We now turn to analyze conditions (\ref{cond_const}) and (\ref{cond_linear_2}).
Since the spectrum of the Ising chain is limited, there is no approximation analog to the Debye
approximation for the harmonic chain and $\Delta_{\mu}^2$
cannot be expressed in terms of $E_{\mu-1}$ and $E_{\mu}$. 
We therefore approximate (\ref{cond_const}) and (\ref{cond_linear_2}) with simpler expressions.
The results are thus quantitatively not as precise as for the harmonic chain, but nevertheless
yield reliable order of magnitute estimates.

Let us first analyze condition (\ref{cond_const}). Since it cannot be checked for
every state $\ve a \ket$ we use the stronger condition
\begin{equation}
\label{cond_const_ising}
E_{\mu} - \frac{E_0}{N_G} > \beta \, \left[\Delta_{\mu}^2 \right]_{max},
\end{equation}
instead, which implies that (\ref{cond_const}) holds for all states $\ve a \ket$.
We require (\ref{cond_const_ising}) to be true for all states with energies in the range (\ref{e_range}).
It is hardest to satisfy for $E_{\mu} = E_{min}$, we thus get the condition on $n$:
\begin{equation}
\label{cond_const_ising2}
n > \beta \, \frac{\left[\Delta_{\mu}^2 \right]_{max}}{e_{min} - e_0},
\end{equation}
where $e_{min} = E_{min} / n$ and $e_0 = E_0 / (n N_G)$.

We now turn to analyze condition (\ref{cond_linear_2}).
Equation (\ref{ising_delta_a}) shows that the $\Delta_{\mu}^2$ do not contain terms which are
proportional to $E_{\mu}$. One thus has to determine, when the $\Delta_{\mu}^2$ are approximately constant
which is the case if
\begin{equation}
\label{min_max_linear}
\beta \, \frac{\left[ \Delta_{\mu}^2 \right]_{max} - \left[ \Delta_{\mu}^2 \right]_{min}}{2} \ll
\left[ E_{\mu} \right]_{max} - \left[ E_{\mu} \right]_{min},
\end{equation}
where $[ x ]_{max}$ and $[ x ]_{min}$ denote the maximal and minimal value $x$ takes on in all states
$\ve a \ket$. As a direct consequence, we get
\begin{equation}
\left| c_1 \right| \ll 1
\end{equation}
which means that temperature is intensive. Defining the quantity $e_{\mu} = E_{\mu} / n$, we can rewrite
(\ref{min_max_linear}) as a condition on $n$,
\begin{equation}
\label{min_max_linear2}
n \ge
\frac{\beta}{2 \, \delta} \, \frac{\left[ \Delta_{\mu}^2 \right]_{max} - \left[ \Delta_{\mu}^2 \right]_{min}}
{\left[ e_{\mu} \right]_{max} - \left[ e_{\mu} \right]_{min}}
\end{equation}
where the accuracy parameter $\delta \ll 1$ is equal to the ratio of the lhs and the rhs of
(\ref{min_max_linear}).

Since equation (\ref{min_max_linear}) does not take into account the energy range (\ref{e_range}),
its application needs some further discussion.

If the occupation number of one mode of a group is changed, say from $n_k^a(\mu) = 0$ to $n_k^a(\mu) = 1$,
the corresponding $\Delta_{\mu}^2$ differ at most by $4 \, B^2\, \left| K^2 - L^2 \right| \, / \, (n+1)$.
On the other hand,
$\left[ \Delta_{\mu}^2 \right]_{max} - \left[ \Delta_{\mu}^2 \right]_{min} = B^2 \, \left| K^2 - L^2 \right|$.
The state with the maximal $\Delta_{\mu}^2$ and the state with the minimal $\Delta_{\mu}^2$ thus differ
in nearly all occupation numbers and therefore their difference in energy is close to
$\left[ E_{\mu} \right]_{max} - \left[ E_{\mu} \right]_{min}$. On the other hand,
states with similar energies $E_{\mu}$ also have a similar $\Delta_{\mu}^2$.
Hence the $\Delta_{\mu}^2$ only change quasi continuously with energy and equation (\ref{min_max_linear})
ensures that the $\Delta_{\mu}^2$ are approximately constant even on only a part of the possible energy range.

We are now going to discuss three special coupling models.

%
%-----------------------------------------------------------------------------------------------------------
%-----------------------------------------------------------------------------------------------------------
%
\subsection{Coupling with constant width $\Delta_a$: $J_y = 0$}

If one of the couplings vanishes ($J_x = 0$ or $J_y = 0$), $K = L$ and $\Delta_{\mu}^2 = B^2 \, K^2$
is constant.
In this case only criterion (\ref{cond_const}) has to be satisfied,
which then coincides with (\ref{cond_const_ising2}).

Plugging expressions (\ref{ising_e_quer}), (\ref{ising_e_0}) and (\ref{ising_delta_a}) with 
$J_x = J$ and $J_y = 0$ into condition (\ref{cond_const_ising2}), one can now
calculate the minimal number of systems per group.
%
% ---------------------------------------------------------------------------
%
% Figure K=L
%
\begin{figure}[h]
\psfrag{-8.1}{\small \raisebox{-0.1cm}{$10^{-8}$}}
\psfrag{-6.1}{\small \raisebox{-0.1cm}{$10^{-6}$}}
\psfrag{-4.1}{\small \raisebox{-0.1cm}{$10^{-4}$}}
\psfrag{-2.1}{\small \raisebox{-0.1cm}{$10^{-2}$}}
\psfrag{2.1}{\small \raisebox{-0.1cm}{$10^{2}$}}
\psfrag{2}{\small \hspace{+0.2cm} $10^{2}$}
\psfrag{4}{\small \hspace{+0.2cm} $10^{4}$}
\psfrag{6}{\small \hspace{+0.2cm} $10^{6}$}
\psfrag{8}{\small \hspace{+0.2cm} $10^{8}$}
\psfrag{n}{\raisebox{0.1cm}{$n_{min}$}}
\psfrag{c1}{$\: T / B$}
\epsfig{file=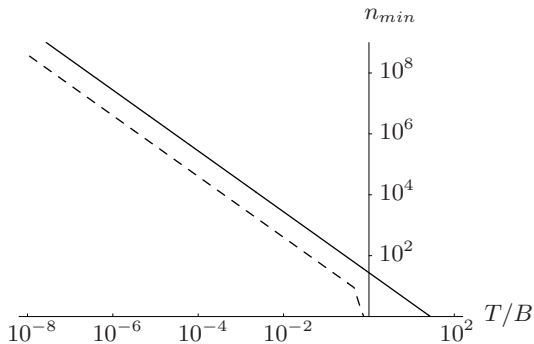,width=7cm}
\caption{Log-log-plot of $n_{min}$ from eq. (\ref{cond_const_ising2}) for $K = L = 0.1$ (dashed line) and
for $K = L = 10$ (solid line) as a function of $T / B$. $\alpha = 10$ is defined in equation
(\ref{e_range}).}
\label{K=L}
\end{figure}

Figure \ref{K=L} shows $n_{min}$ for weak coupling $K = L = 0.1$ and strong coupling
$K = L = 10$ with $\alpha = 10$ as a function of $T / B$.
We choose units where Boltzmann's constant $k_B$ is one.

For any set of parameters, there is a finite temperature above which $n_{min} = 1$.

Note that, since $\Delta_{\mu} = const$, condition (\ref{cond_const_ising2}) coincides with
criterion (\ref{cond_const}) ($\Delta_{\mu} = const = \left[ \Delta_{\mu} \right]_{max}$),
so that using (\ref{cond_const_ising2}) does not involve any approximations.

\begin{figure}[t]
\psfrag{-0.51}{\tiny \raisebox{-0.1cm}{$10^{-0.5}$}}
\psfrag{0.51}{\tiny \raisebox{-0.1cm}{$10^{0.5}$}}
\psfrag{1.21}{\tiny \raisebox{-0.1cm}{$10^{1.2}$}}
\psfrag{1.41}{\tiny \raisebox{-0.1cm}{$10^{1.4}$}}
\psfrag{0.2}{\tiny \hspace{-0.4cm} $10^{0.2}$}
\psfrag{0.4}{\tiny \hspace{-0.4cm} $10^{0.4}$}
\psfrag{0.5}{\tiny \hspace{-0.4cm} $10^{0.5}$}
\psfrag{1}{\tiny \hspace{-0.55cm} $10^{1.0}$}
\psfrag{1.5}{\tiny \hspace{+0.2cm} $ $}
\psfrag{k}{\hspace{-1cm} \raisebox{+0.3cm}{{\small $K=L=0.1$}}}
\psfrag{l}{\hspace{-0.4cm} {\small $K=L=10$}}
\psfrag{a1}{\small \hspace{-0.6cm} \raisebox{-0.1cm}{$\alpha=1$}}
\psfrag{a2}{\small \hspace{-0.45cm} $\alpha=100$}
\psfrag{b1}{\small \hspace{-0.6cm} \raisebox{-0.2cm}{$\alpha=1$}}
\psfrag{b2}{\small \hspace{-0.2cm} $\alpha=100$}
\psfrag{n}{\raisebox{0.1cm}{$n_{min}$}}
\psfrag{c1}{$\, \frac{T}{B}$}
\epsfig{file=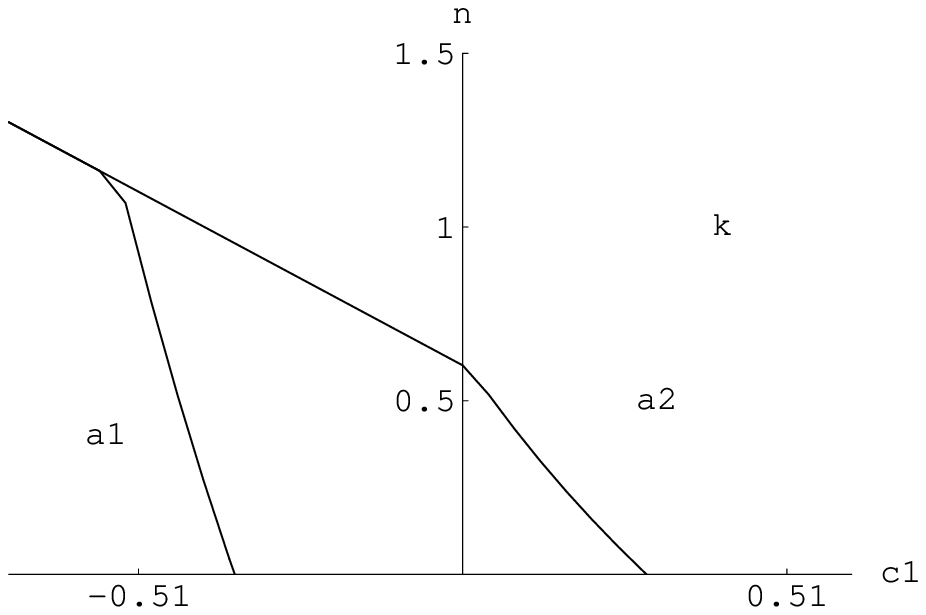,width=3.8cm}
\hspace{0.4cm}
\epsfig{file=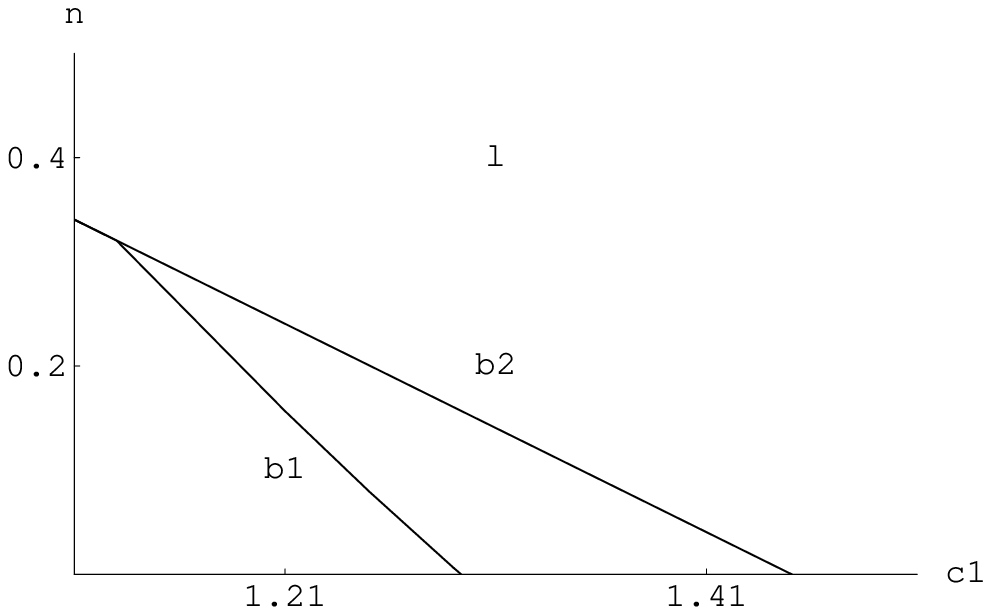,width=3.8cm}
\caption{Log-log-plot of $n_{min}$ as a function of $T / B$ from eq. (\ref{cond_const_ising2})
for two values of the accuracy parameter $\alpha$, $\alpha = 1$ and $\alpha = 100$,
left plot for $K = L = 0.1$ and right plot for $K = L = 10$.
$\alpha$ is defined in equation (\ref{e_range}).}
\label{alphadep}
\end{figure}

As condition (\ref{cond_linear_1}) is automatically satisfied for the present model,
the results do not depend on the ``accuracy parameter'' $\delta$.
The dependence of the results on $\alpha$ is shown in figure \ref{alphadep}. $\alpha$ plays a role only where
$E_{min}= \overline{E} / (\alpha N_G) + E_0 / N_G$ (cf. eq. (\ref{e_range})). Then for smaller $\alpha$,
$n_{min}$ eventually decays steeper and thus reaches $n_{min} = 1$ already at lower
temperatures. There is thus a temperature interval, where $n_{min}$ is larger for larger $\alpha$
and vice versa. This dependency has the same interpretation as for the harmonic chain.
%
% ---------------------------------------------------------------------------
%
% Figure alphadep
%
%

%
%-----------------------------------------------------------------------------------------------------------
%-----------------------------------------------------------------------------------------------------------
%
\subsection{Fully anisotropic coupling: $J_x = -J_y$}

If both couplings are nonzero, the variances $\Delta_{\mu}^2$ are not constant.
As an example, we consider here the
fully anisotropic coupling, where $J_x = - J_y$, i. e. $K = 0$.
Now criteria (\ref{cond_const_ising2}) and (\ref{min_max_linear2}) have to be met.

For $K = 0$, one has $\left[ \Delta_{\mu}^2 \right]_{max} = B^2 \, L^2$,
$\left[ \Delta_{\mu}^2 \right]_{min} = 0$ and
$\left[ e_{\mu} \right]_{max} = - \left[ e_{\mu} \right]_{min} = B$.

Plugging these results into (\ref{min_max_linear2}) as well as (\ref{ising_e_quer}) and (\ref{ising_e_0}) 
into (\ref{cond_const_ising2}), the minimal number of systems per group can be calculated.

\begin{figure}[h]
\psfrag{-6.1}{\small \raisebox{-0.1cm}{$10^{-6}$}}
\psfrag{-4.1}{\small \raisebox{-0.1cm}{$10^{-4}$}}
\psfrag{-2.1}{\small \raisebox{-0.1cm}{$10^{-2}$}}
\psfrag{2.1}{\small \raisebox{-0.1cm}{$10^{2}$}}
\psfrag{4.1}{\small \raisebox{-0.1cm}{$10^{4}$}}
\psfrag{2}{\small \hspace{+0.2cm} $10^{2}$}
\psfrag{4}{\small \hspace{+0.2cm} $10^{4}$}
\psfrag{6}{\small \hspace{+0.2cm} $10^{6}$}
\psfrag{8}{\small \hspace{+0.2cm} $10^{8}$}
\psfrag{n}{\raisebox{0.1cm}{$n_{min}$}}
\psfrag{c1}{$\: T / B$}
\epsfig{file=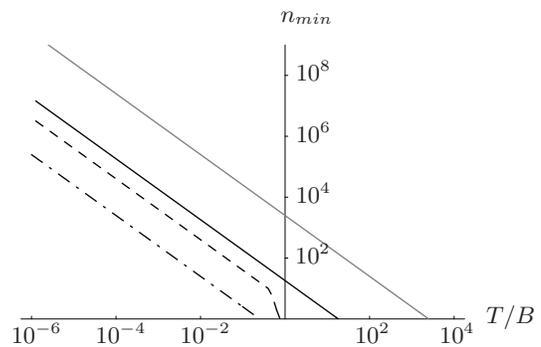,width=7cm}
\caption{Log-log-plot of $n_{min}$ for $L = 0.1$ from eq. (\ref{cond_const_ising2}) (dashed line) and
from eq. (\ref{min_max_linear2}) (dash - dotted line) and $n_{min}$ for $L = 10$
from eq. (\ref{cond_const_ising2}) (solid line) and from eq. (\ref{min_max_linear2}) (gray line)
as a function of $T / B$. $K = 0$, $\alpha = 10$ and $\delta = 0.01$.
$\alpha$ and $\delta$ are defined in equations (\ref{e_range}) and (\ref{min_max_linear2}) respectively.}
\label{K=0}
\end{figure}

Figure \ref{K=0} shows $n_{min}$ from criterion (\ref{cond_const_ising2}) and
from criterion (\ref{min_max_linear2}) separately, for weak coupling $L = 0.1$ and strong coupling
$L = 10$ with $\alpha = 10$ and $\delta = 0.01$ as a function of $T / B$.
For each coupling strength $L$, the stronger condition, that is the higher curve in figure \ref{K=0}, sets
the relevant lower bound to the group size $n$. 

%
% ---------------------------------------------------------------------------
%
% Figure K=0
%
In the present case, all occupation numbers $n_k^a(\mu)$ are zero in the ground state of a group.
In this state, $\Delta_{\mu}^2$ is maximal ($\Delta_{\mu}^2 = B^2 \, L^2$) as can be seen from
(\ref{ising_delta_a}). Therefore criterion (\ref{cond_const_ising2}) is equivalent to
criterion (\ref{cond_const}) for low temperatures, where $E_{min} = \left[ E_{\mu} \right]_{min}$.
For high temperatures, where $E_{min} = \overline{E} / (\alpha N_G)$, condition (\ref{cond_const_ising2})
is slightly stronger than (\ref{cond_const}). For the present model, this is only the case for
$L = 0.1$ (dashed line) and $T \gtrsim 0.45 B$.

In figure \ref{K=0}, the results obtained from equation (\ref{min_max_linear2}) are proportional to
$\delta^{-1}$ (dash - dotted line and gray line), while those obtained from equation (\ref{cond_const_ising2})
(dashed line and solid line) have the same dependency on $\alpha$ as shown in figure \ref{alphadep}.

%
%-----------------------------------------------------------------------------------------------------------
%-----------------------------------------------------------------------------------------------------------
%
\subsection{Isotropic coupling: $J_x = J_y$}

As a third example, we consider the isotropic coupling,
where $J_x = J_y$, i. e. $L = 0$.
Again, both criteria (\ref{cond_const_ising2}) and (\ref{min_max_linear2}) have to be met.

The values of $\left[ \Delta_{\mu}^2 \right]_{max}$,
$\left[ \Delta_{\mu}^2 \right]_{min}$, $\left[ e_{\mu} \right]_{max}$ and
$\left[ e_{\mu} \right]_{min}$ are given in equations (\ref{e_min_max_k}), (\ref{e_min_max_K})
and (\ref{delta_min_max})

For the present model with $L = 0$ and $|K| < 1$ all occupation numbers $n_k^a(\mu)$ are zero in the
ground state and thus $\Delta_{\mu}^2 = 0$. As a consequence, condition (\ref{cond_const_ising2})
cannot be used instead of (\ref{cond_const}). We therefore argue as follows:
In the ground state $E_{\mu} - E_0 / N_G = 0$ as well as $\Delta_{\mu}^2 = 0$ and all
occupation numbers $n_k^a(\mu)$ are zero. If one occupation number is then changed from $0$ to $1$,
$\Delta_{\mu}^2$ changes at most by $4 \, B^2 \, K^2 / (n+1)$ and $E_{\mu}$ changes at least by
$2 \, B \, ( 1 - |K| )$. Therefore (\ref{cond_const}) will hold for all states except the ground state
if
\begin{equation}\label{special_cond_const}
n > 2 \, B \, \beta \, \frac{K^2}{1 - |K|}
\end{equation}

If $|K| > 1$, occupation numbers of modes with $\cos (k) < 1 / |K|$ are zero in the ground state
and occupation numbers of modes with $\cos (k) > 1 / |K|$ are one.
$\Delta_{\mu}^2$ for the ground state then is
$\left[ \Delta_{\mu}^2 \right]_{gs} \approx \left[ \Delta_{\mu}^2 \right]_{max} / 2$ and
(\ref{cond_const_ising2}) is a good approximation of condition (\ref{cond_const}).

Plugging these results into (\ref{min_max_linear2}) as well as (\ref{ising_e_quer}) and (\ref{ising_e_0})
into (\ref{cond_const_ising2}) for $|K| > 1$ and using (\ref{special_cond_const}) for $|K| < 1$,
the minimal number of systems per group can be calculated.

\begin{figure}[h]
\psfrag{-6.1}{\small \raisebox{-0.1cm}{$10^{-6}$}}
\psfrag{-4.1}{\small \raisebox{-0.1cm}{$10^{-4}$}}
\psfrag{-2.1}{\small \raisebox{-0.1cm}{$10^{-2}$}}
\psfrag{2.1}{\small \raisebox{-0.1cm}{$10^{2}$}}
\psfrag{4.1}{\small \raisebox{-0.1cm}{$10^{4}$}}
\psfrag{2}{\small \hspace{+0.2cm} $10^{2}$}
\psfrag{4}{\small \hspace{+0.2cm} $10^{4}$}
\psfrag{6}{\small \hspace{+0.2cm} $10^{6}$}
\psfrag{8}{\small \hspace{+0.2cm} $10^{8}$}
\psfrag{10}{\small \hspace{+0.35cm} $10^{10}$}
\psfrag{12}{\small \hspace{+0.35cm} $10^{12}$}
\psfrag{n}{\raisebox{0.1cm}{$n_{min}$}}
\psfrag{c1}{$\: T / B$}
\epsfig{file=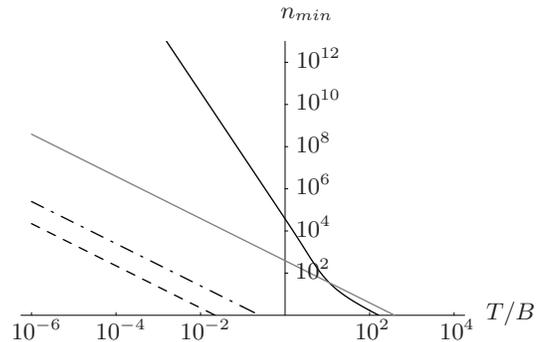,width=7cm}
\caption{Log-log-plot of $n_{min}$ for $K = 0.1$ from eq. (\ref{special_cond_const}) (dashed line),
and from eq. (\ref{min_max_linear2}) (dash - dotted line) and $n_{min}$ for $K = 10$
from eq. (\ref{cond_const_ising2}) (solid line) and from eq. (\ref{min_max_linear2}) (gray line)
as a function of $T / B$. $L = 0$, $\alpha = 10$ and
$\delta = 0.01$. $\alpha$ and $\delta$ are defined in equations (\ref{e_range}) and (\ref{min_max_linear2})
respectively.}
\label{L=0}
\end{figure}

Figure \ref{L=0} shows $n_{min}$ from criteria (\ref{special_cond_const}) and (\ref{min_max_linear2})
for weak coupling $K = 0.1$ and from criteria (\ref{cond_const_ising2}) and (\ref{min_max_linear2})
for strong coupling $K = 10$ with $\alpha = 10$ and $\delta = 0.01$ as a function of $T / B$.
For each coupling strength $K$, the stronger condition, that is the higher curve in figure \ref{L=0}, sets
the relevant lower bound to the group size $n$. 

%
% ---------------------------------------------------------------------------
%
% Figure L=0
%

Equation (\ref{special_cond_const}) does not take into account the relevant energy range (\ref{e_range}),
it is therefore possible that a weaker condition could be sufficient in that case. However, since
(\ref{min_max_linear2}) is a stronger condition than (\ref{special_cond_const}) for $K = 0.1$,
this possibility has no relevance.

For strong coupling, $K = 10$ (\ref{cond_const_ising2}) is used to approximate (\ref{cond_const}).
This approximation is expected to be good because $\Delta_{\mu}$ is close to its maximal value for low energy
states. Furthermore, the temperature dependence we obtain here for $n_{min}$ for low temperatures
is the same as for the harmonic chain, $n_{min} \propto T^{-3}$. This agreement is to be expected:
The two couplings, when expressed in creation and annihilation operators, have the same structure and the upper
limit of the spectrum of the spin chain becomes irrelevant at low temperatures.   

For the present model, the dependence of the results on the ``accuracy parameters'' $\alpha$ and $\delta$
is as follows. Results obtained from equation (\ref{min_max_linear2}) are proportional to
$\delta^{-1}$ (dash - dotted line and gray line), while the result obtained from equation
(\ref{cond_const_ising2})
(solid line) has the same dependency on $\alpha$ as shown in figure \ref{alphadep}. For weak coupling
and low temperatures (dashed line) $n_{min}$ does not depend on the two ``accuracy parameters''.

%-------------------------------------------------------------------------------------------------------
%XXXXXXXXXXXXXXXXXXXXXXXXXXXXXXXXXXXXXXXXXXXXXXXXXXXXXXXXXXXXXXXXXXXXXXXXXXXXXXXXXXXXXXXXXXXXXXXXXXXXXXX
%XXXXXXXXXXXXXXXXXXXXXXXXXXXXXXXXXXXXXXXXXXXXXXXXXXXXXXXXXXXXXXXXXXXXXXXXXXXXXXXXXXXXXXXXXXXXXXXXXXXXXXX
%-------------------------------------------------------------------------------------------------------
%
\section{Summary and Conclusions\label{conclusion}}

We have considered a linear chain of particles interacting with their nearest neighbors.
We have partitioned the chain into identical groups of $n$ adjoining particles each.
Taking the number of such groups to be very large and assuming the total
system to be in a thermal state with temperature $T$ we have found conditions
(equations (\ref{cond_const}) and (\ref{cond_linear_2})), which ensure that each group is approximately
in a thermal state. Furthermore, we have determined when the isolated groups have the same temperature $T$,
that is when temperature is intensive.

The result shows that, in the quantum regime,
these conditions depend on the temperature $T$, contrary to the classical case.
The characteristics of the temperature dependence are determined by the width $\Delta_a$ of the distribution
of the total energy eigenvalues in a product state and its dependence on the group energies $E_a$.
The low temperature behavior, in particular, is related to the fact that $\Delta_a$ has a nonzero
minimal value. This fact does not only appear in the harmonic chain or spin chains 
but is a general feature of quantum systems composed of interacting particles or subsystems.
The commutator $[H,H_0]$ is nonzero and the ground state of the total
system is energetically lower than the lowest product state, therefore $\Delta_a$ is nonzero, even
at zero temperature \cite{Wang2002,Jordan2003,Allahverdyan2002,Nieuwenhuizen2002}.

We have then applied the general method to a harmonic chain and several types of Ising spin chains.
For concrete models, the conditions (\ref{cond_const}) and (\ref{cond_linear_2}) determine a 
minimal group size and thus a minimal length scale
on which temperature may be defined according to the temperature concept we adopt.
Grains of size below this length scale are no more in a thermal state. Thus temperature measurements
with a higher resolution should no longer be interpreted in a standard way.

We have given order of magnitude estimates for the minimal group size (minimal length scale) for
the models mentioned above.
The most striking difference between the spin chains and the harmonic chain is that the energy spectrum
of the spin chains is limited, while it is infinite for the harmonic chain.

For spins at very high global temperatures, the total density matrix is then almost completely mixed, i. e.
proportional to the identity matrix, and thus does not change under basis transformations.
There are thus global temperatures which are high enough, so that local temperatures exist even for
single spins.

For the harmonic chain, this feature does not appear, since the size of the relevant energy range
increases indefinitely with growing global temperature, leading to the constant minimal length
scale in the high energy range.

For the spin chain with isotropic coupling, $J_x = J_y$, and the harmonic chain,
the temperature dependencies of $n_{min}$
for low temperatures coincide, $n_{min} \propto T^{-3}$, because both couplings have the
same structure and the upper limit of the spectrum of the spin chain becomes irrelevant at
low temperatures.   

The spin chain with $J_x = 0$ or $J_y = 0$ shows the interesting feature that $\Delta_a^2$ is constant and
condition (\ref{cond_linear_2}) is automatically fulfilled.

The set of models we have discussed is by no means exhaustive. It would be particularly interesting to
see whether there are systems for which local temperatures can exist although they are not intensive.
This can happen if either $\varepsilon_a$ or $\Delta_a^2$ were proportional to $E_a$.
$\Delta_a^2$ however has dimension energy squared, so that it cannot be proportional to $E_a$
unless there exists another characteristic energy of the system independent of $E_a$.
So far, we have not found models where $\varepsilon_a \propto E_a$.

For the models we consider here, the off diagonal elements of the density operator in
the product basis, $\bra a \ve \hat \rho \ve b \ket$ ($a \not= b$), are
significantly smaller than the diagonal ones, $\bra a \ve \hat \rho \ve a \ket$.
Our general result, conditions (\ref{cond_const}) and (\ref{cond_linear_2}), thus
states that the density matrix $\hat \rho$ ``approximately'' factorizes with respect to
the considered partition. This implies that the state $\hat \rho$ is not entangled
with respect to this partition, at least within the chosen accuracy.
It would therefore be interesting to see how our result relates to
the scaling of entanglement in many particle systems \cite{Vidal2003}.

Unfortunately, our approach only applies to nonzero temperatures. The underlying central limit theorem
\cite{Hartmann2003,Hartmann2004b} is about the weak convergence of the distribution of energy eigenvalues.
Weak convergence means that only integrals over energy intervals of nonzero length do converge.
We thus cannot make statements about a system in its groundstate let alone about
the entanglement in that state.

Since harmonic lattice models in Debye approximation have proven to be successful
in modeling thermal properties of insulators (e.g. heat capacity) \cite{Kittel1983},
our calculation for the harmonic chain provides a first estimate of the minimal length scale on which intensive
temperatures exist in insulating solids,
\begin{equation}
\label{length}
l_{min} = n_{min} \, a_0.
\end{equation}
Let us give some numerical estimates: Choosing the ``accuracy parameters''
to be $\alpha = 10$ and $\delta = 0.01$,
we get for hot iron ($T \gg \Theta \approx 470 \,$K, $a_0 \approx 2.5 \,${\AA})
$l_{min} \approx 50 \,\mu$m, while
for carbon ($\Theta \approx 2230 \,$K, $a_0 \approx 1.5 \,${\AA}) at room temperature ($270 \,$K)
$l_{min} \approx 10 \,\mu$m. The coarse-graining will experimentally be most relevant at very low
temperatures,
where $l_{min}$ may even become macroscopic. A pertinent example is silicon
($\Theta \approx 645 \,$K, $a_0 \approx 2.4 \,${\AA}),
which has $l_{min} \approx 10 \,$cm at $T \approx 1 \,$K
(again with $\alpha = 10$ and $\delta = 0.01$).

Of course the validity of the harmonic lattice model
will eventually break down at finit, high temperatures and our estimates will thus no longer apply there.

Measurable consequences of the local breakdown of the concept of temperature and their
implications for future nanotechnology are interesting questions which arise in the
context of the present discussion.

In the secnarios of global equilibrium, which we consider here,
a temperature measurement with a microscopic thermometer, locally in thermal contact with
the large chain, would not reveal the non existence of local temperature.
One can model such a measurement with a small system, representing the thermometer,
coupled to a heat bath, representing the chain. It is a known result of such system bath
models \cite{Weiss1999},
that the system always relaxes to a thermal state with the global temperature
of the bath, no matter how local the coupling might be.

This, however, does not mean that the existence or non existence of local temperatures
had no physical relevance: There are indeed physical properties, which are determined
by the local states rather than the global ones. Whether these properties are of
thermal character depends on the existence of local temperatures.
A detailed discussion of such properties will be given elsewhere. 

The length scales, calculated in this paper, should also constrain the way one can meaningfully
define temperature profiles in non-equilibrium scenarios \cite{Michel2003}. Here,
temperature measurements with a microscopic thermometer, which is locally in thermal
contact with the sample, might indeed be suitable to measure the local temperature.
An explicit study of this possibility should be subject of future research.

We thank M.\ Michel, M.\ Henrich, H.\ Schmidt, M.\ Stollsteimer, F.\ Tonner and C.\ Kostoglou
for fruitful discussions.

%
%-----------------------------------------------------------------------------------------------
%XXXXXXXXXXXXXXXXXXXXXXXXXXXXXXXXXXXXXXXXXXXXXXXXXXXXXXXXXXXXXXXXXXXXXXXXXXXXXXXXXXXXXXXXXXXXXXX
%XXXXXXXXXXXXXXXXXXXXXXXXXXXXXXXXXXXXXXXXXXXXXXXXXXXXXXXXXXXXXXXXXXXXXXXXXXXXXXXXXXXXXXXXXXXXXXX
%-----------------------------------------------------------------------------------------------
%
\appendix
\section{Diagonalization of the Harmonic Chain \label{diagon_harmonic_chain}}

The Hamiltonian of a harmonic chain is diagonalized by a Fourier transformation
and the definition of creation and annihilation operators.

For the entire chain with periodic boundary conditions, the Fourier transformation reads
\begin{equation}
\left\{
\begin{array}{c}
q_{j}\\ p_{j}
\end{array} \right\} =
\frac{1}{\sqrt{n N_G}}
\sum_{k} 
\left\{ \begin{array}{c}
u_{k} \exp (i a_0 k j) \\ v_{k} \exp (- i a_0 k j) 
\end{array} \right\}
\end{equation}
with $k = 2 \pi l / (a_0 \, n \, N_G)$ and $(l = 0, \pm 1, \dots,$ \linebreak $\pm (n N_G - 2) / 2,
\, (n N_G) / 2$,
where $n N_G$ has been assumed to be even.

For the diagonalization of one single group, the Fourier transformation is
\begin{equation}
\left\{
\begin{array}{c}
q_{j}\\ p_{j}
\end{array} \right\} =
\sqrt{\frac{2}{n+1}}
\sum_{k} 
\left\{ \begin{array}{c}
u_{k}\\ v_{k}
\end{array} \right\}
\sin (a_0 k j)
\end{equation}
with
$k = \pi l / (a_0 \, (n+1))$ and $(l = 1, 2, \dots, n)$.

The definition of the creation and annihilation operators is in both cases
\begin{equation}
\left\{
\begin{array}{c}
a_{k}^{\dagger} \\ a_{k}
\end{array} \right\} =
\frac{1}{\sqrt{2 m \omega_{k}}} \:
\left( m \omega_{k} u_{k}
\left\{
\begin{array}{c}
- \\ +
\end{array} \right\}
i v_{k} \right) 
\end{equation}
where the corresponding $u_{k}$ and $v_{k}$ have to be inserted.
The frequencies $\omega_{k}$ are given by $\omega^2_{k} = 4 \omega_0^2 \sin^2(k a_0 / 2)$ in both cases..

The operators $a_{k}^{\dagger}$ and $a_{k}$ satisfy bosonic commutation relations
\begin{eqnarray}
[a_{k}, a_{p} ] & = & 0 \nn \\
\left[\right.a_{k}, a_{p}^{\dagger} \left.\right] & = & \delta_{k p}
\end{eqnarray}
and the diagonalized Hamiltonian reads
\begin{equation}
H = \sum_k \omega_k \left(a_{k}^{\dagger} a_k + \frac{1}{2} \right)
\end{equation}
%
%
%-----------------------------------------------------------------------------------------------
%XXXXXXXXXXXXXXXXXXXXXXXXXXXXXXXXXXXXXXXXXXXXXXXXXXXXXXXXXXXXXXXXXXXXXXXXXXXXXXXXXXXXXXXXXXXXXXX
%XXXXXXXXXXXXXXXXXXXXXXXXXXXXXXXXXXXXXXXXXXXXXXXXXXXXXXXXXXXXXXXXXXXXXXXXXXXXXXXXXXXXXXXXXXXXXXX
%-----------------------------------------------------------------------------------------------
%
%
\section{Diagonalization of the Ising Chain \label{diagon_ising_chain}}

The Hamiltonian of the Ising chain is diagonalized via Jordan-Wigner transformation which maps it
to a fermionic system \cite{Katsura1962,Lieb1961}.
\begin{eqnarray}
c_i & = & \left( \prod_{j < i} \sigma_j^z \right) \frac{\sigma_i^x + i \sigma_i^y}{2} \nn \\
c_i^{\dagger} & = & \left( \prod_{j < i} \sigma_j^z \right) \frac{\sigma_i^x - i \sigma_i^y}{2}
\end{eqnarray}
The operators $c_i$ and $c_i^{\dagger}$ fulfill fermionic anti-commutation relations
\begin{eqnarray}
\label{fermionic_comm}
\{c_i, c_j \} & = & 0 \nn \\
\{c_i, c_j^{\dagger} \} & = & \delta_{i j}
\end{eqnarray}
and the Hamiltonian reads 
\begin{eqnarray} \label{fermionic_ising_ham}
H & = & B \left[ \sum_j \left(2 c_j^{\dagger} c_j - 1 \right) -
K \sum_j \left( c_j^{\dagger} c_{j+1} + \textrm{h.c.} \right) - \right. \nn \\
& - & \left. L \sum_j \left( c_j^{\dagger} c_{j+1}^{\dagger} + \textrm{h.c.} \right) \right]
\end{eqnarray}
with $K = (J_x + J_y) / (2 B)$ and $L = (J_x - J_y) / (2 B)$.
In the case of periodic boundary conditions a boundary term is neglected in equation
(\ref{fermionic_ising_ham}).
For long chains ($n N_G \rightarrow \infty$) this term is suppressed by a factor $(n N_G)^{-1}$.
The Hamiltonian now describes Fermions which interact with their nearest neighbors.
As for the bosonic system, a Fourier transformations maps the system to noninteracting fermions.
For the whole chain with periodic boundary conditions
\begin{equation}
\left\{
\begin{array}{c}
c_j^{\dagger} \\ c_j
\end{array} \right\}
= \frac{1}{\sqrt{n N_G}} \sum_k e^{i k j}
\left\{
\begin{array}{c}
d_k^{\dagger} \\ d_k
\end{array} \right\}
\end{equation}
with $k = (2 \pi l) / (n N_G)$ where
$l = 0, \pm 1, \dots,$ \linebreak $\pm (n N_G - 2) / 2, \, (n N_G) / 2$ for $n N_G$ even,
and
\begin{equation}
\left\{
\begin{array}{c}
c_j^{\dagger} \\ c_j
\end{array} \right\}
= \sqrt{\frac{2}{n + 1}} \sum_k \sin (k j)
\left\{
\begin{array}{c}
d_k^{\dagger} \\ d_k
\end{array} \right\}
\end{equation}
with $k = (\pi l) / (n + 1)$ and ($l = 1, 2, \dots, n$) for one single group.

In the case of periodic boundary conditions, fermion interactions of the form $d_k^{\dagger} d_{-k}^{\dagger}$
and $d_k d_{-k}$ remain.
Therefore, one still has to apply a Bogoliubov transformation to diagonalize the system, i.e.
\begin{eqnarray}
d_k^{\dagger} & = & u_k b_k^{\dagger} - i v_k b_{-k} \nn \\
d_k & = & u_k b_k + i v_k b_{-k}^{\dagger}
\end{eqnarray}
where $u_k = u_{-k}$, $v_k = - v_{-k}$ and $u_k^2 + v_k^2 = 1$.
With the definitions $u_k = \cos(\Theta_k / 2)$ and $v_k = \sin(\Theta_k / 2)$ the interaction
terms disappear for
\begin{equation}
\cos(\Theta_k) =
\frac{1 - K \cos k}{\sqrt{[1 - K \cos k]^2 + [L \sin k]^2}}
\end{equation}

In the case of the finite chain of one group, the  Bogoliubov transformation is not needed since
the corresponding terms are of the form $d_k^{\dagger} d_k^{\dagger}$ and $d_k d_k$ and vanish by
virtue of equation (\ref{fermionic_comm}).

The Hamiltonians in the diagonal form read
\begin{equation}
H = \sum_k \omega_k \left( b_k^{\dagger} b_k - \frac{1}{2} \right)
\end{equation}
where the frequencies are
\begin{equation} \label{ising_frequ}
\omega_k = 2 B \sqrt{[1 - K \cos k]^2 + [L \sin k]^2}
\end{equation}
with $k = (2 \pi l) / (n N_G)$ for the periodic chain and
\begin{equation}
\omega_k = 2 B \left( 1 - K \cos k \right)
\end{equation}
with $k = (\pi l) / (n + 1)$ for the finite chain.

For the finite chain the occupation number operators may also
be chosen such that $\omega_k$ is always positive. Here, the convention at hand is more convenient,
since the same occupation numbers also appear in the group interaction and thus in $\Delta_{\mu}$.

\subsection{Maxima and minima of $E_{\mu}$ and $\Delta_{\mu}^2$}

The maximal and minimal values of $E_{\mu}$ are given by
\begin{equation}\label{e_min_max_k}
\left\{ \begin{array}{c} \left[E_{\mu} \right]_{max} \\ \left[E_{\mu}\right]_{min} \end{array} \right\} =
\left\{ \begin{array}{c} + \\ - \end{array} \right\} \, n \, B,
\end{equation}
for $| K | < 1$ and by
\begin{align}\label{e_min_max_K}
& \left\{ \begin{array}{c} \left[E_{\mu} \right]_{max} \\ \left[E_{\mu}\right]_{min} \end{array} \right\} =
\nn \\
& =
\left\{ \begin{array}{c} + \\ - \end{array} \right\} \, n \, B \,
\frac{2}{\pi} \left[ \sqrt{K^2 - 1} + \arcsin \left( \frac{1}{| K |} \right) \right],
\end{align}
for $| K | > 1$, where the sum over all modes $k$ has been approximated with an integral.

The maximal and minimal values of $\Delta_{\mu}^2$ are given by
\begin{equation}\label{delta_min_max}
\left\{ \begin{array}{c} \left[ \Delta_{\mu}^2 \right]_{max}
\\ \left[ \Delta_{\mu}^2 \right]_{min} \end{array} \right\}
= B^2 \left\{ \begin{array}{c} \textrm{max} \left( K^2 , L^2 \right) \\
\textrm{min} \left( K^2 , L^2 \right) \end{array} \right\}.
\end{equation}
%

%
%
%-----------------------------------------------------------------------------------------------
%XXXXXXXXXXXXXXXXXXXXXXXXXXXXXXXXXXXXXXXXXXXXXXXXXXXXXXXXXXXXXXXXXXXXXXXXXXXXXXXXXXXXXXXXXXXXXXX
%XXXXXXXXXXXXXXXXXXXXXXXXXXXXXXXXXXXXXXXXXXXXXXXXXXXXXXXXXXXXXXXXXXXXXXXXXXXXXXXXXXXXXXXXXXXXXXX
%-----------------------------------------------------------------------------------------------
%

\end{document}